\documentclass[aps,prl,twocolumn,showpacs,superscriptaddress]{revtex4-1}
\usepackage[dvips,pdftex]{graphicx}
\usepackage{hyperref}
\usepackage{amsmath}
\usepackage{ulem}
\usepackage{color}
\usepackage[dvipsnames]{xcolor}



\begin{document}
\title{Quantum subdiffusion with two- and three-body interactions}

\author{I.I.~Yusipov}
\affiliation{Department of Applied Mathematics, Lobachevsky State University of Nizhny Novgorod, Russia}

\author{T.V.~Laptyeva}
\affiliation{Theory of Control and Dynamical Systems Department, Lobachevsky State University of Nizhny Novgorod, Russia}

\author{A.Yu.~Pirova}
\affiliation{Department of Mathematical Software and Supercomputing Technologies, Lobachevsky State University of Nizhny Novgorod, Russia}

\author{I.B.~Meyerov}
\affiliation{Department of Mathematical Software and Supercomputing Technologies, Lobachevsky State University of Nizhny Novgorod, Russia}

\author{S.~Flach}
\affiliation{Center for Theoretical Physics of Complex Systems, Institute for Basic Science, Daejeon 305-732, Korea}

\author{M.V.~Ivanchenko}
\email[]{ivanchenko.mv@gmail.com}
\affiliation{Department of Applied Mathematics, Lobachevsky State University of Nizhny Novgorod, Russia}

\date{\today}

\begin{abstract}
We study the dynamics of a few-quantum-particle cloud in the presence of two- and three-body interactions in weakly disordered one-dimensional lattices. The interaction is dramatically enhancing the Anderson localization length $\xi_1$ of noninteracting particles. We launch compact wave packets and show that
few-body interactions lead to transient subdiffusion of wave packets, $m_2 \sim t^{\alpha}$, $\alpha<1$, on length scales beyond $\xi_1$. The subdiffusion exponent is independent of the
number of particles. Two-body interactions yield $\alpha\approx0.5$ for two and three particles, while three-body interactions decrease it to $\alpha\approx0.2$. The tails of expanding wave packets exhibit exponential localization with a slowly decreasing exponent. We relate our results to subdiffusion in nonlinear random lattices, and to results on restricted diffusion in high-dimensional spaces 
like e.g. on comb lattices.  
\end{abstract}


\maketitle

\section{Introduction}

Localization of quantum and classical waves due to random (Anderson) or quasiperiodic (Aubry-Andre) spatial inhomogeneity of an underlying lattice potential is a fundamental physical phenomenon manifested by light, sound, and matter waves \cite{Anderson,Aubry,Kramers,Evers,Experiments} (see also a recent review \cite{our_review}). To date, rigorous results are available for the case of non-interacting quantum particles (or classical linear waves) only, while the many-body localization problem remains a complex open field, despite some recent advances \cite{Basko,Aleiner,Shlyapnikov}. 

The computationally accessible case of few interacting quantum particles is therefore highly important and can give insights into the above complexity. Earlier studies of two interacting particles (2IP) in random lattices concluded that interaction inflates the single particle localization length $\xi_1$ up to another finite localization length $\xi_2$ \cite{Dorohov,Shep94,Imry,Roemer,Frahm,Pichard1999,Krimer11,Frahm2016}. For quasiperiodic potentials the early predictions differed from an incremental increase of localization length to a decrease of localization length in the insulating regime \cite{Shep_96,Shreib,Evan}.

Recent results reveal much more dramatic and unexpected effects of interactions. First, it was shown that some of the {2IP} states in quasiperiodic lattices can become completely delocalized
in the presence of a non-perturbatively strong interaction, giving rise to an unconstrained wave packet propagation \cite{Khomeriki2012,Frahm2015}. Second, in disordered lattices, it was found that 2IP produce self-sustained subdiffusive propagation beyond the single particle localization length, provided that the disorder is weak and $\xi_2\gg\xi_1$ \cite{Ivanchenko2014}. This regime was associated with quantum chaos and high effective connectivity of states due to interaction \cite{Ivanchenko2014,Krimer2015,Frahm2016}.    

These findings intriguingly correlate with results obtained in the mean field approximation for (infinitely) many interacting particles, which lead to Gross-Pitaevsky type nonlinear wave equations \cite{Dalfovo}. There nonlinearity destroys localization through deterministic chaos and leads again to subdiffusive wave packet propagation \cite{molina,Pikovsky_Shepelyansky,Flach09,laptyeva10,bodyfelt11,Pikovsky,larcher12,Skokos2014}. The particular footprints of the chaotic nature of subdiffusion are seen in an asymptotically self-similar expansion of the wave packet \cite{nde} and a subdiffusion exponent which depends on the power of nonlinearity \cite{nonlinearity_HO}, which is the classical analogue of defining the number of quantum particles
needed for interaction, like two-body or three-body interactions.
  
In this paper we demonstrate that few-body interactions lead to subdiffusive propagation beyond the single-particle localization length for 2IP and three interacting particles (3IP). 
We will denote the specific cases with three particles as 3IP$_k$ where $k=2$ holds for three particles with two-body interactions, and $k=3$ for three particles and exclusive three-body interactions.
In the latter case a new interaction-induced localization length scale 
$\xi_3 \gg \xi_1$ will set the spreading limits for wave packets.
While two-body interactions are the typical situation encountered, three-body interactions appear to be more exotic. However, even such cases are currently experimentally addressable e.g. with
cold polar molecules in optical lattices driven by microwave fields \cite{buechler2007}.

We launch compact wave packets and show that
few-body interactions lead to transient subdiffusion of wave packets, $m_2 \sim t^{\alpha}$, $\alpha<1$, on length scales beyond $\xi_1$. The subdiffusion exponent is not depending on the
number of particles. Two-body interactions yield $\alpha\approx0.5$ for two and three particles, while three-body interactions decrease it to $\alpha\approx0.2$. The tails of expanding wave packets exhibit exponential localization with a slowly decreasing exponent. We relate our results to subdiffusion in nonlinear random lattices, and to results on restricted diffusion in high-dimensional spaces 
like e.g. on comb lattices.  

\begin{figure}[ht!!!]
\begin{center}
(a)\includegraphics[angle=0,width=0.9\columnwidth,keepaspectratio,clip]{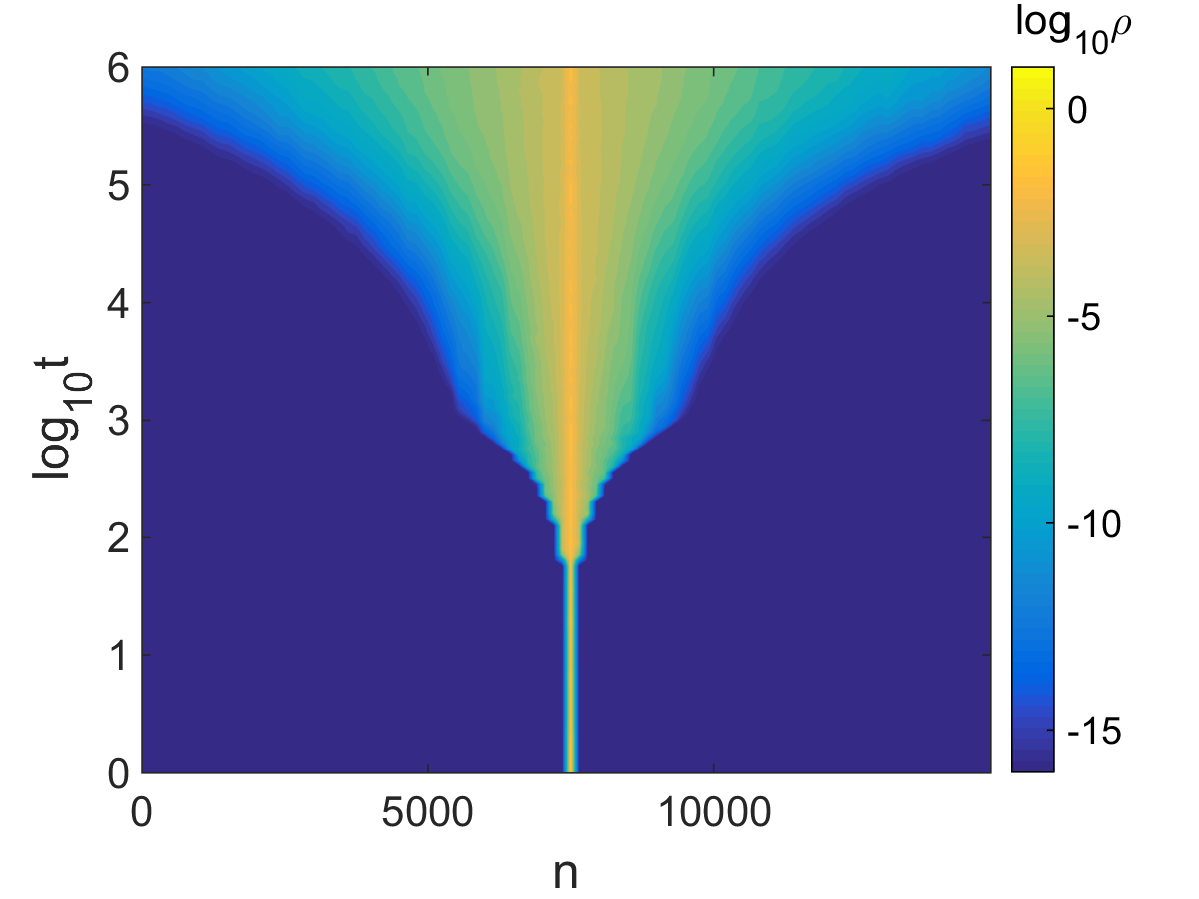}
(b)\includegraphics[angle=0,width=0.9\columnwidth,keepaspectratio,clip]{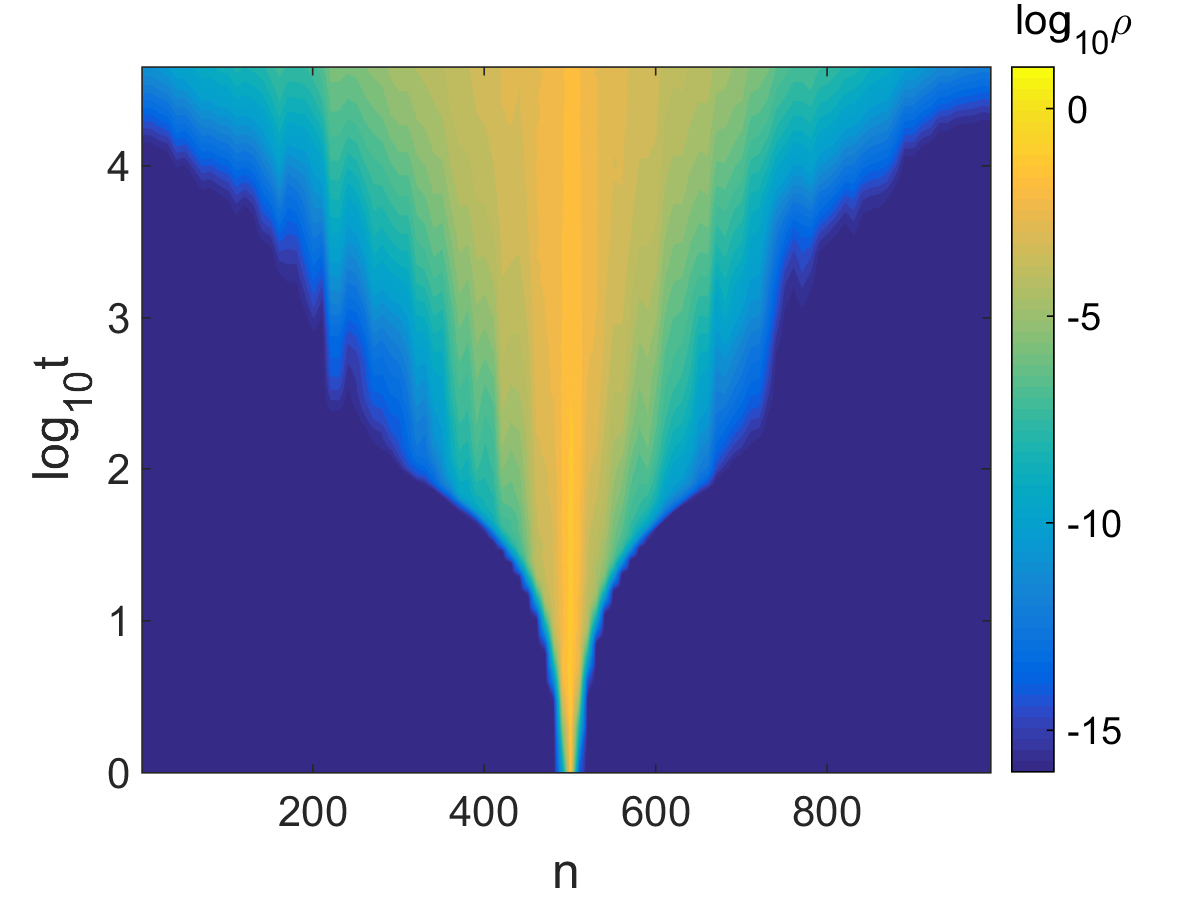}
(c)\includegraphics[angle=0,width=0.9\columnwidth,keepaspectratio,clip]{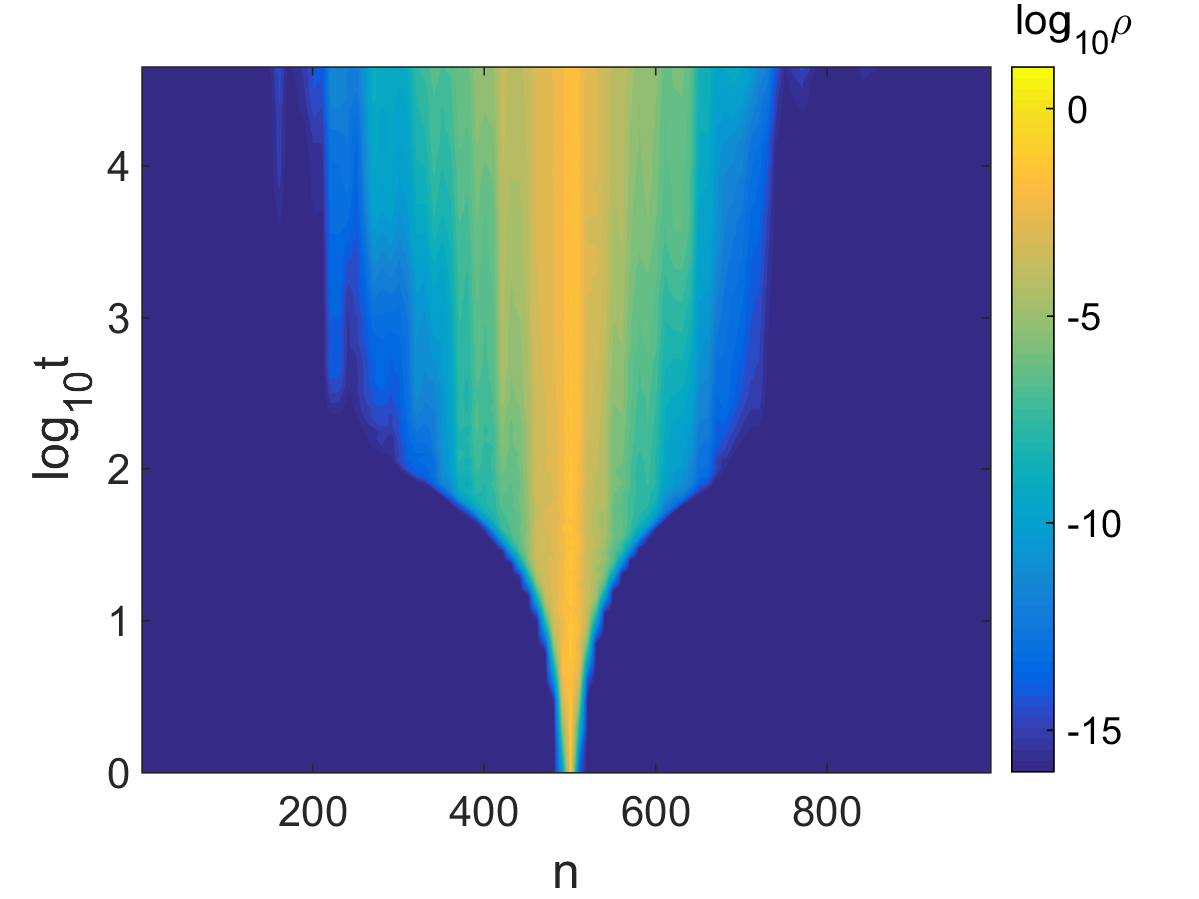}
\caption{
Evolution of ${\rho_n(t)}$ for expanding wave packets on a disordered lattice for (a) 2IP, $W=1.0, U_2=2.0, N=15000$, 
(b) $\mbox{3IP}_2$, $W=3.0, U_2=2.0, U_3=0, N=1000$, and (c) $\mbox{3IP}_3$, $W=3.0, U_2=0.0, U_3=2.0, N=1000$. 
Note the crossover from ballistic to subdiffusive expansion which takes place at about $t=10^3$, once the wave packet reaches the size $\xi_1$, especially clearly visible in (a). 
{ 
The ballistic expansion is ongoing from $t=0$, and the seeming constant width of the central part for $1 < t < 100$ in (a) is due to color code discretization (compare with Fig.\ref{fig:2}(a)). 
}
Averaging is taken over $30$ disorder realizations.}
\label{fig:1}
\end{center}
\end{figure}

\section{Model}
We use the Bose-Hubbard model with the Hamiltonian
\begin{equation}
\label{eq1}
{\cal \hat H}= \sum\limits_j\left[\hat b_{j+1}^+\hat b_j+\hat
b_{j}^+\hat b_{j+1}+\epsilon_j\hat b_{j}^+\hat b_j+\frac{U_k}{k!}
\left(\hat b_{j}^+\right)^k\left(\hat b_{j}\right)^k \right]
\end{equation}
where $\hat b_{j}^+$ and $\hat b_{j}$ are creation and
annihilation operators of indistinguishable bosons at lattice site
$j$. $U_k$ measures the $k$-body on-site interaction strength.
The on-site energies are random uncorrelated numbers with a uniform probability density function on the interval $\epsilon_j\in[-W/2,W/2]$.

We make use of the basis $|j,l\rangle\equiv\hat b_j^+b_l^+|0\rangle$ for 2IP and $|j,l,m\rangle\equiv\hat b_j^+b_l^+b_m^+|0\rangle$ for 3IP, where $|0\rangle $ is the vacuum state. The 2IP wave function $\Psi=\sum\limits_{j,l}\varphi_{j,l}|j,l\rangle$ evolves according to the Schr\"odinger equation
\begin{equation}
\label{eq2}
i\dot{\varphi}_{j,l}=\epsilon_{j,l}\varphi_{j,l}+\sum\limits_\pm(\varphi_{j\pm1,l}+\varphi_{j,l\pm1}),
\end{equation}
where $\epsilon_{j,l}=\epsilon_j+\epsilon_l+U_2\delta_{j,l}$ and $\delta_{j,l}$ is the Kronecker symbol.

For the 3IP case the corresponding equations read:
\begin{equation}
\label{eq2a}
i\dot{\varphi}_{j,l,m}=\epsilon_{j,l,m}\varphi_{j,l,m}+\sum\limits_\pm(\varphi_{j\pm1,l,m}+\varphi_{j,l\pm1,m}+\varphi_{j,l,m\pm1}),
\end{equation}
where  $\epsilon_{j,l,m}=\epsilon_j+\epsilon_l+\epsilon_m+U_2\left(\delta_{j,l}+\delta_{l,m}+\delta_{j,m}\right)$ for 3IP$_2$ 
and $\epsilon_{j,l,m}=\epsilon_j+\epsilon_l+\epsilon_m+U_3\delta_{j,l}\delta_{l,m}$ for 3IP$_3$.

In the absence of interactions, $U_k=0$, the solutions to (\ref{eq2}) and (\ref{eq2a}) reduce to tensor products of single particle (SP) eigenstates of
\begin{equation}
\label{eq3}
i\dot{\varphi}_{m}=\epsilon_{m}\varphi_{m}+\varphi_{m+1}+\varphi_{m-1}.
\end{equation}
The SP eigenstates { $\varphi_m(t) = A_m^{(p)} {\rm e}^{-i\lambda_p t}$} of (\ref{eq3}) with energy $-2-W/2< \lambda_p <2+W/2$ are exponentially localized: $A_{|m| \rightarrow \infty}^{(p)} \sim 
{\rm e}^{-|m|/\xi_1(\lambda_p)}$ with a maximal localization length well-estimated by $\xi_1\approx 96/W^2$ for $W < 4$, \cite{Kramers}. 

\begin{figure*}
\begin{center}
(a)\includegraphics[angle=0,width=0.9\columnwidth,keepaspectratio,clip]{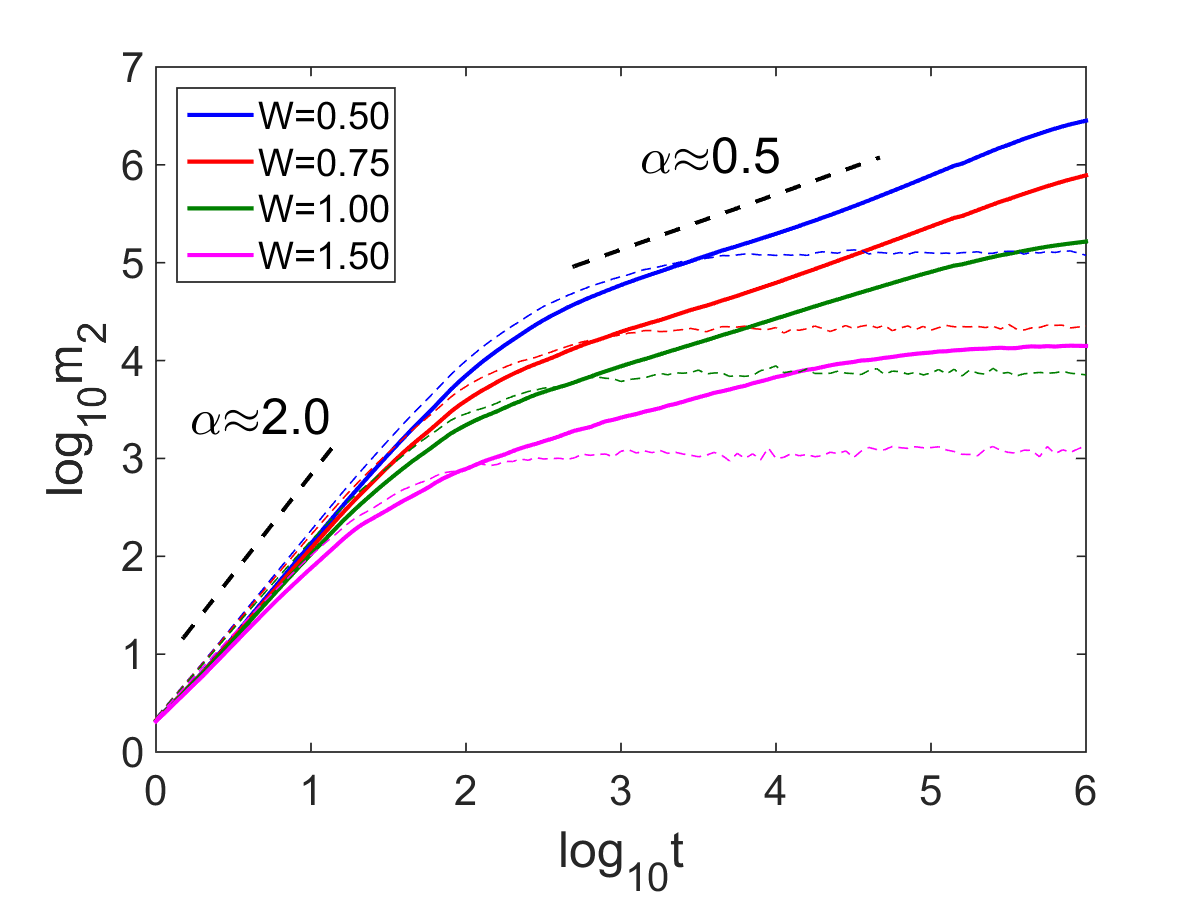}
(b)\includegraphics[width=0.9\columnwidth,keepaspectratio,clip]{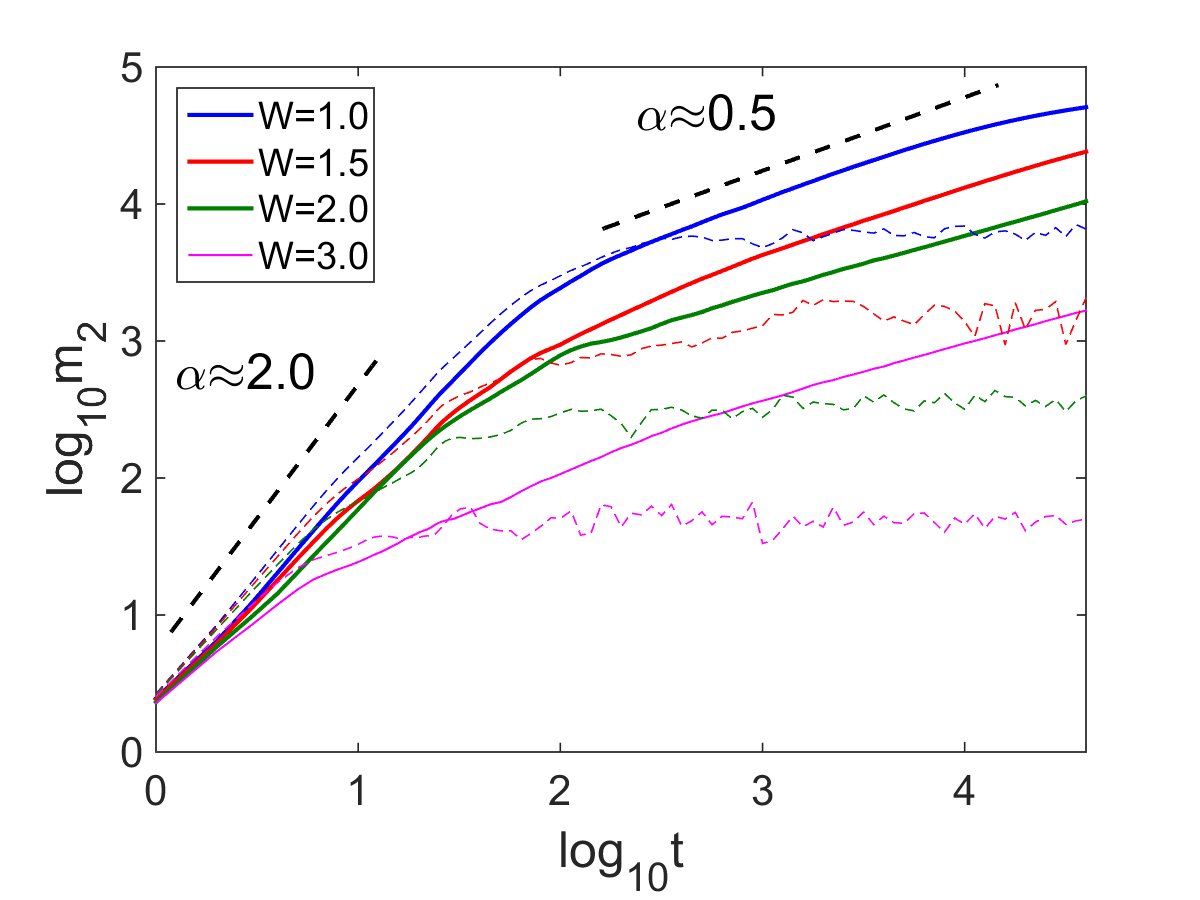}
(c)\includegraphics[angle=0,width=0.9\columnwidth,keepaspectratio,clip]{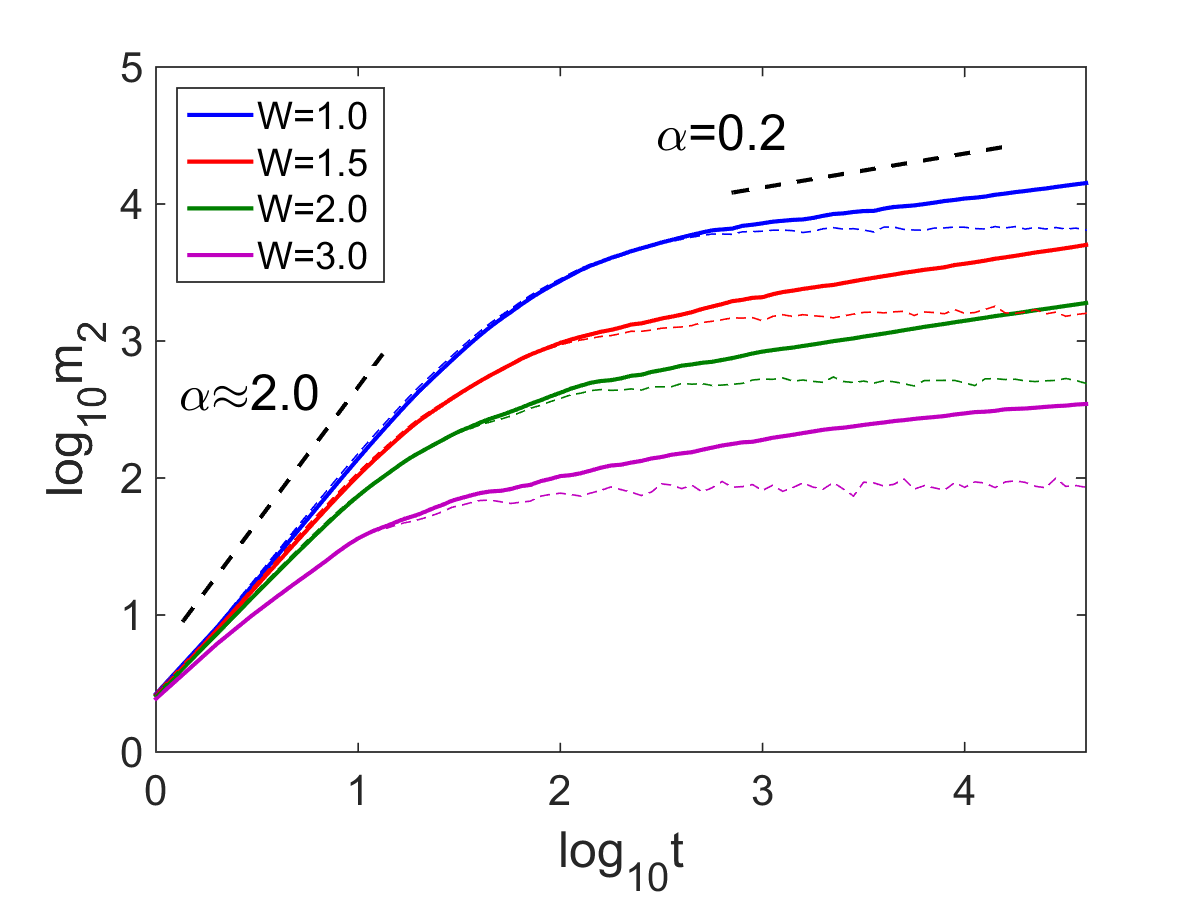}
(d)\includegraphics[angle=0,width=0.9\columnwidth,keepaspectratio,clip]{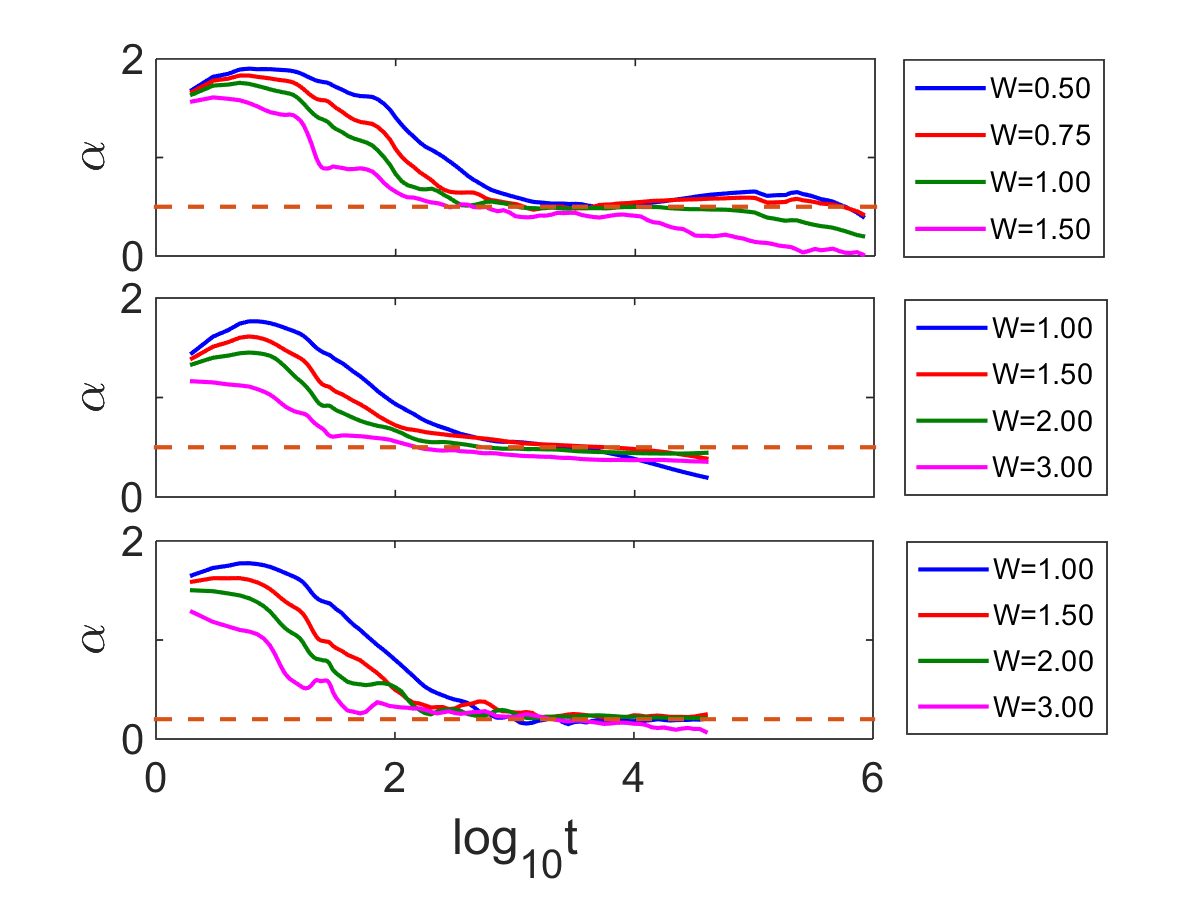}
\caption{Time-dependence of the wave packet second moment,$m_2(t)$ for (a) 2IP, $U_2=2.0$, $N=10000$, (b) $\mbox{3IP}_2$, $U_2=2.0, U_3=0, N=1000$, and (c) $\mbox{3IP}_3$, $U_2=0.0, U_3=2.0, N=1000$.  
The corresponding cases of vanishing interactions $U_2 = U_3= 0$  are shown with dashed lines. Thick black dashed straight lines guide the eye for algebraic growth $m_2\sim t^\alpha$, $\alpha = 2, 0.5, 0.2$. (d) Evolution of the instantaneous subdiffusion exponent, $\alpha = d\log_{10} m_s / d \log_{10} t$, for 2IP (top), ${\rm 3IP}_2$ (middle), and ${\rm 3IP}_3$ (bottom). Horizontal dashed lines correspond to the exponents $0.5$ and $0.2$. Averaging is taken over $30$ disorder realizations. {  Results for 2IP corroborate earlier findings on shorter time scales in \cite{Ivanchenko2014}}. 
}
\label{fig:2}
\end{center}
\end{figure*}

We integrate (\ref{eq2}) and (\ref{eq2a}) numerically on finite lattices $N\times N$ and $N\times N\times N$ using a three-step symplectic integrator coined the PQ method (see Ref. \cite{bodyfelt11} for details).
The initial conditions correspond to particles placed at adjacent sites. To characterize the wave packet dynamics  we calculate the one-dimensional probability distribution function (PDF) of the particle density as
$\rho_j=\sum\limits_l |\varphi_{j,l}|^2$ for 2IP and $\rho_j=\sum\limits_{l,m} |\varphi_{j,l,m}|^2$ for 3IP.
We further compute the time dependence of the first moment  (center-of-mass) $m_1=\sum_j j \rho_j$ and the second moment $m_2=\sum_j(j-m_1)^2 \rho_j$. {  Note that this quantity may increase due to both coherent and independent few particle diffusion. At the same time, the latter remains limited by Anderson localization, and only the former defines growth and scaling of $m_2$ beyond the SP localization volume (cf. \cite{Pichard1999}), which we also verified.}

The system size is varied within 
the limits $ N=1000 ... 15000$ (2IP) and $N=200 ... 1000$ (3IP), {  whereas the integration time reaches $10^6$ for 2IP (outperforming \cite{Ivanchenko2014} by an order of magnitude) and $10^{4.5}$ for 3IP.} The averaging  is done over $30$ disorder realizations.

\section{Results}

Our numerical results generalize earlier findings made for 2IP \cite{Ivanchenko2014} showing that different spreading regimes occur on two spatial scales (Fig.\ref{fig:1}). The evolution of ${\rho_n(t)}$ clearly demonstrates the initial fast wave packet expansion up to the SP localization length $\xi_1$, followed by a slower spreading into a  larger volume available due to the interaction-induced localization length $\xi_2$. The first stage is invariable since non-interacting dynamics dominates. For the second stage, the 2IP and 3IP$_2$ cases with two-body interactions show the same type of subdiffusion
(Fig.\ref{fig:1}(a) and (b)). However three particles with three-body interactions (${\rm3IP}_3$) drastically slow down the subdiffusive expansion beyond $\xi_1$, as compared to
the 3IP$_2$ case.
We are particularly interested in a regime of weak disorder when
the limiting second spatial scales $\xi_2$ and $\xi_3$ are not reached during our simulations.

We follow the evolution of the wave packet second moment, $m_2(t)$ for 2IP, ${\rm 3IP}_2$ and ${\rm 3IP}_3$ (Fig.\ref{fig:2}(a)-(c)). We observe sub-ballistic expansion into the 
single particle volume of size $\xi_1$. The second stage manifests subdiffusive spreading, $m_2(t)\sim t^\alpha$, with exponent $\alpha\approx0.5$ for 2IP and ${\rm3IP}_2$ (Fig.\ref{fig:2}(a,b)), and $\alpha\approx0.2$ for ${\rm3IP}_3$ (Fig.\ref{fig:2}(c)). The respective non-interacting cases, $U_2=U_3=0$, correspond to the dashed lines and demonstrate the halt of expansion once $\xi_1$ is reached. 
We also calculate the local tangent $\alpha(t) = \frac{d\log_{10} m_s }{ d \log_{10} t}$, in order to  estimate the instantaneous subdiffusion exponent $\alpha(t)$. 
In Fig.\ref{fig:2}(d) we observe that the instantaneous exponent $\alpha(t)$ shows a transient time-independent plateau at the corresponding values $0.2$ ($\rm3IP_3$) and $0.5$ ($\rm3IP_2$ and 2IP), whose width increases with decreasing strength
of disorder.
\begin{figure*}
\begin{center}
(a)\includegraphics[angle=0,width=0.9\columnwidth,keepaspectratio,clip]{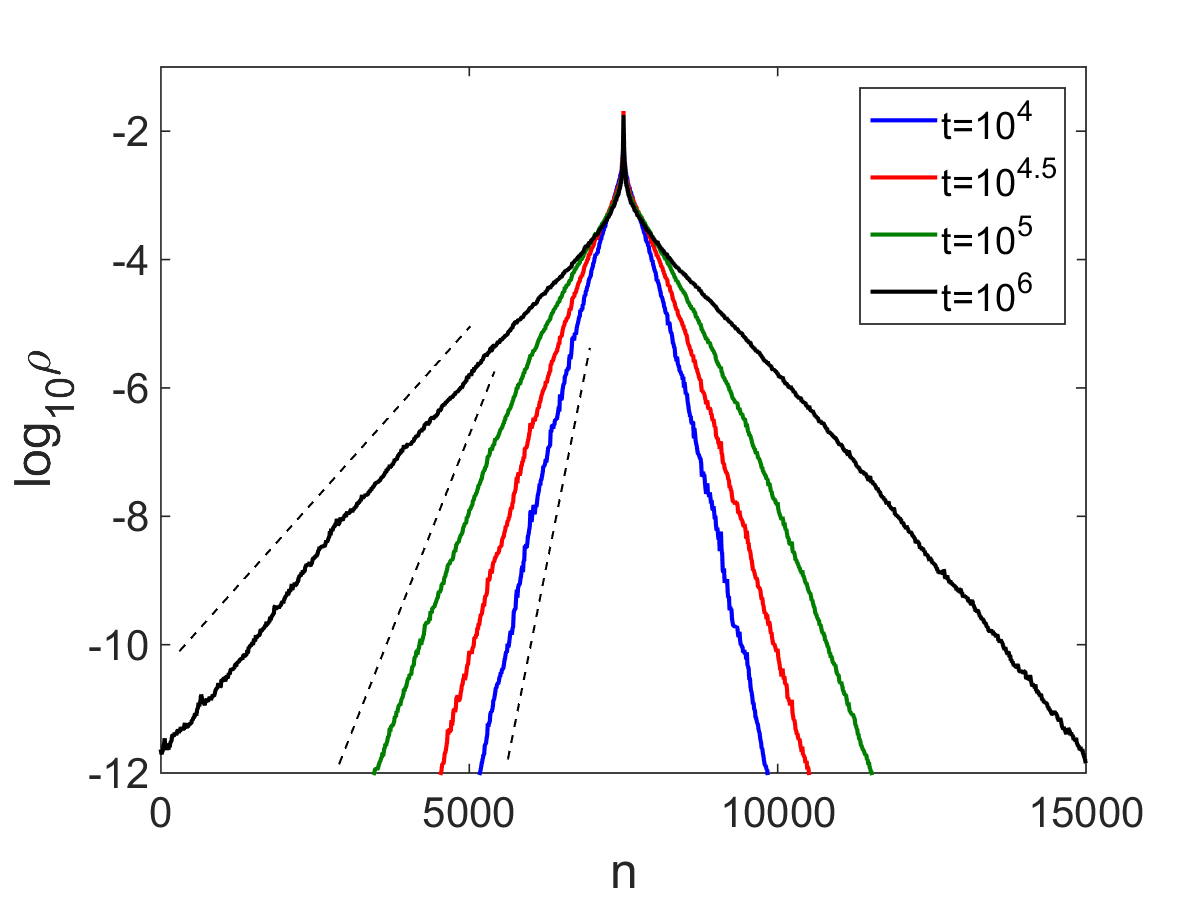}
(b)\includegraphics[width=0.9\columnwidth,keepaspectratio,clip]{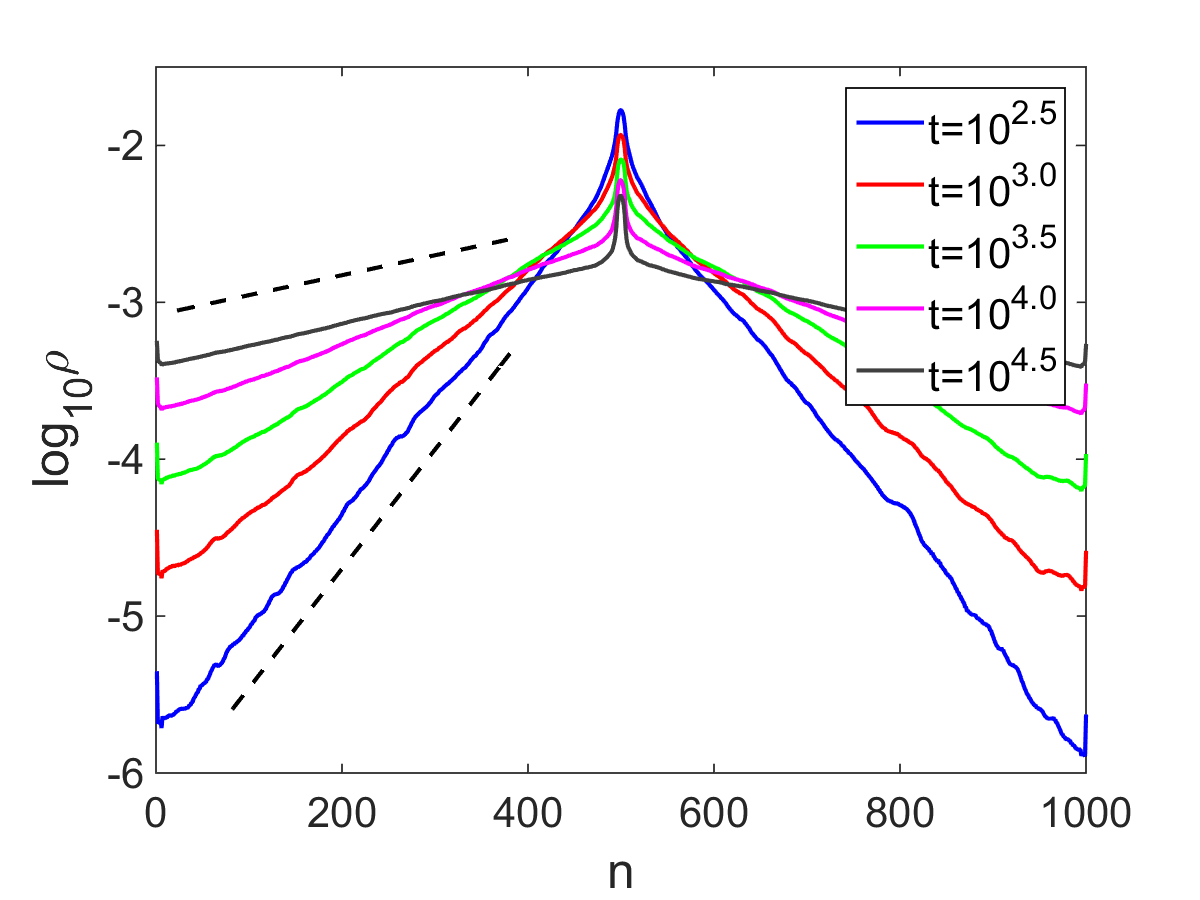}
(c)\includegraphics[angle=0,width=0.9\columnwidth,keepaspectratio,clip]{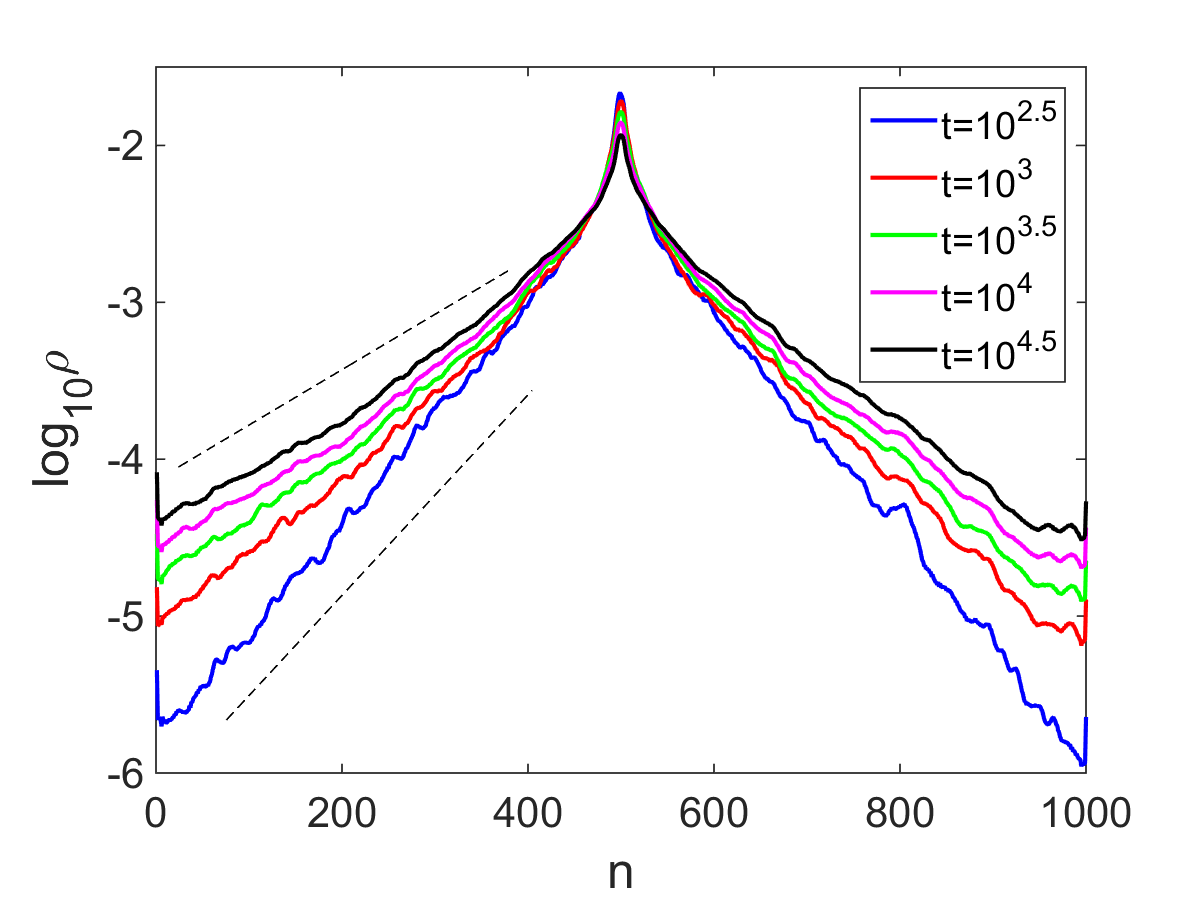}
(d)\includegraphics[angle=0,width=0.9\columnwidth,keepaspectratio,clip]{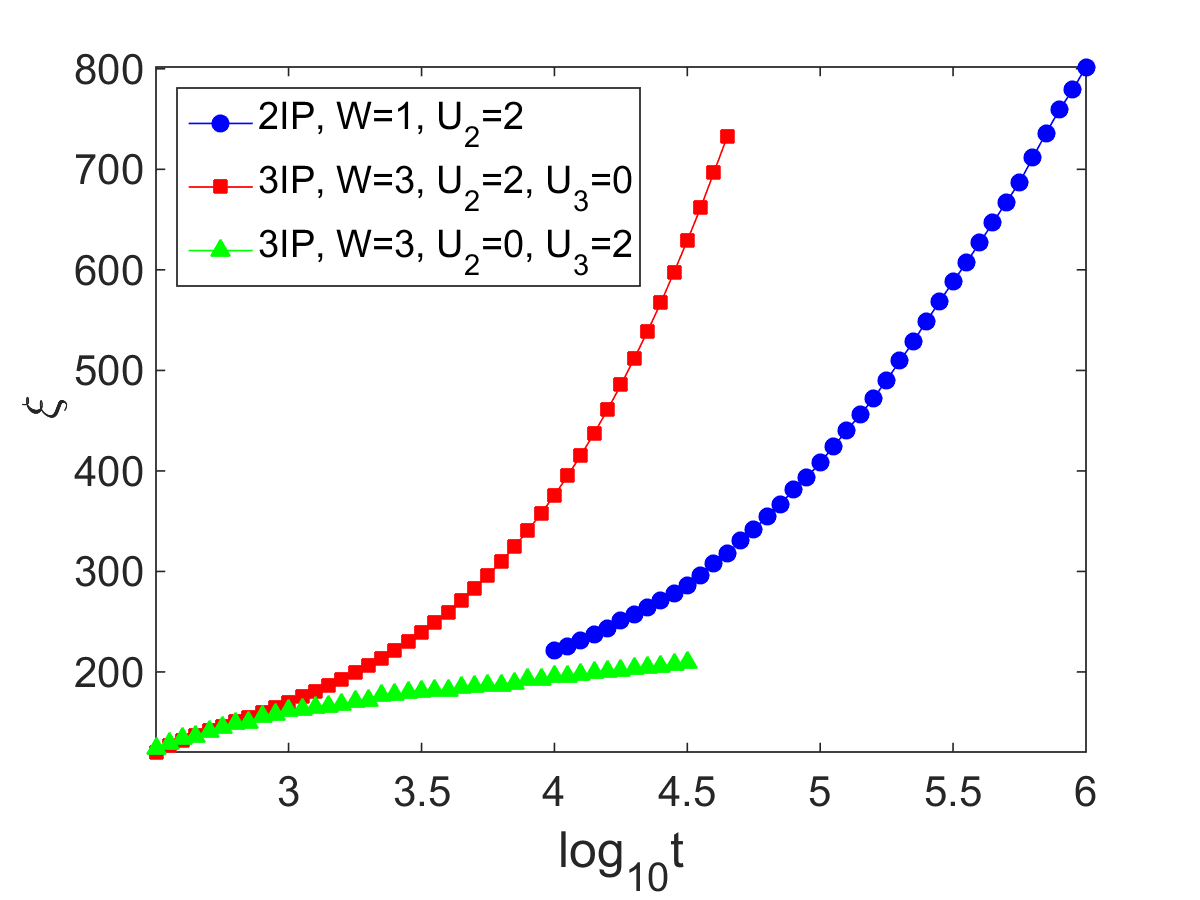}
\caption{Wave packet expansion from a localized initial state on a random lattice for (a) 2IP, $U_2=2.0, N=15000$, (b) $\mbox{3IP}_2$, $U_2=2.0, U_3=0, N=1000$, and (c) $\mbox{3IP}_3$, $U_2=0.0, U_3=2.0, N=1000$. Dashed lines guide the eye to observe slowly varying exponential localization in tails, ${\rho}_l \sim \exp[-(l-l_0)/\xi(t)]$. (d) Slow growth of the dynamical localization length as read from the wave packet tails, $\xi(t)$, for the cases in (a)-(c). Averaging is taken over $30$ disorder realizations. 
}
\label{fig:3}
\end{center}
\end{figure*}

We further address the shape of the spreading wave packet. Fig.~\ref{fig:3}(a)-(c) illustrates the evolution from a localized initial state through snapshots at different times. It shows that in all cases the wave packet profile is well approximated by a decaying exponent with the time-dependent localization length:
\begin{equation}
\label{self}
{\rho}_l\sim \exp[-|l-l_0|/\xi(t)].
\end{equation}
As shown in Fig.\ref{fig:3}(d), the dynamical localization length increases monotonously in time.

\begin{figure*}[ht!!!]
\begin{center}
(a)\includegraphics[angle=-90,width=0.9\columnwidth,keepaspectratio,clip]{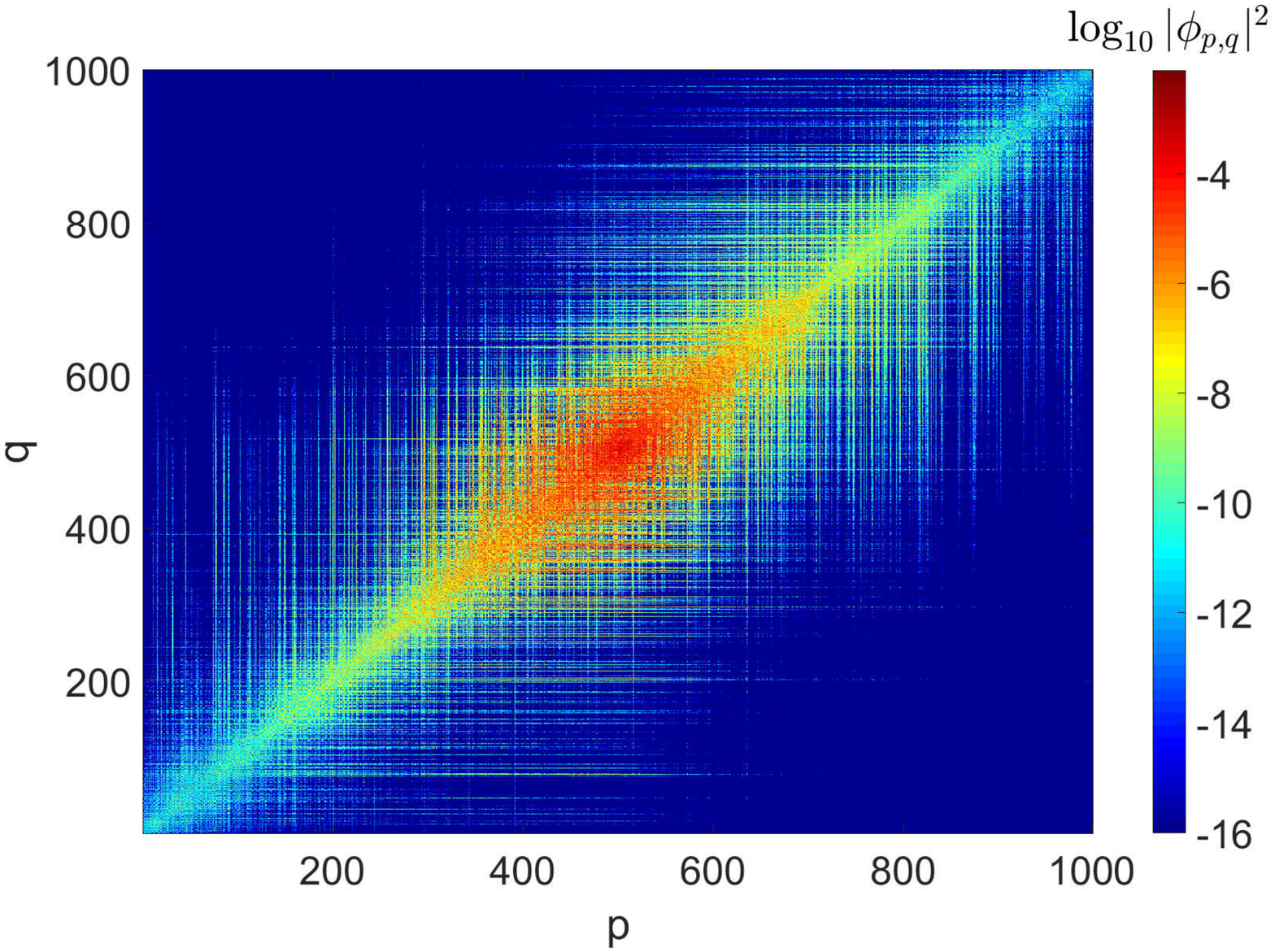}
(b)\includegraphics[angle=-90,width=0.9\columnwidth,keepaspectratio,clip]{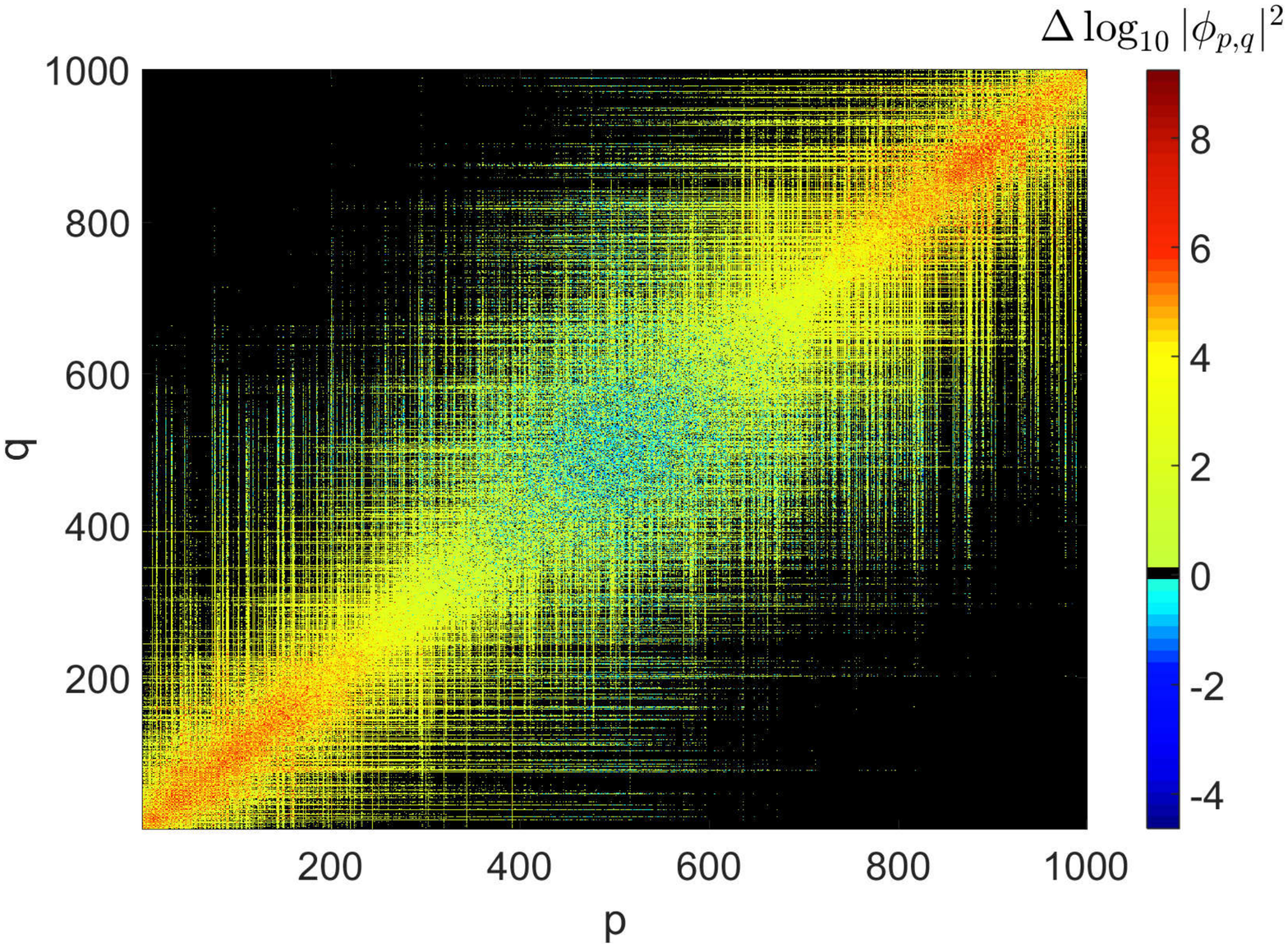}
(c)\includegraphics[angle=-90,width=0.9\columnwidth,keepaspectratio,clip]{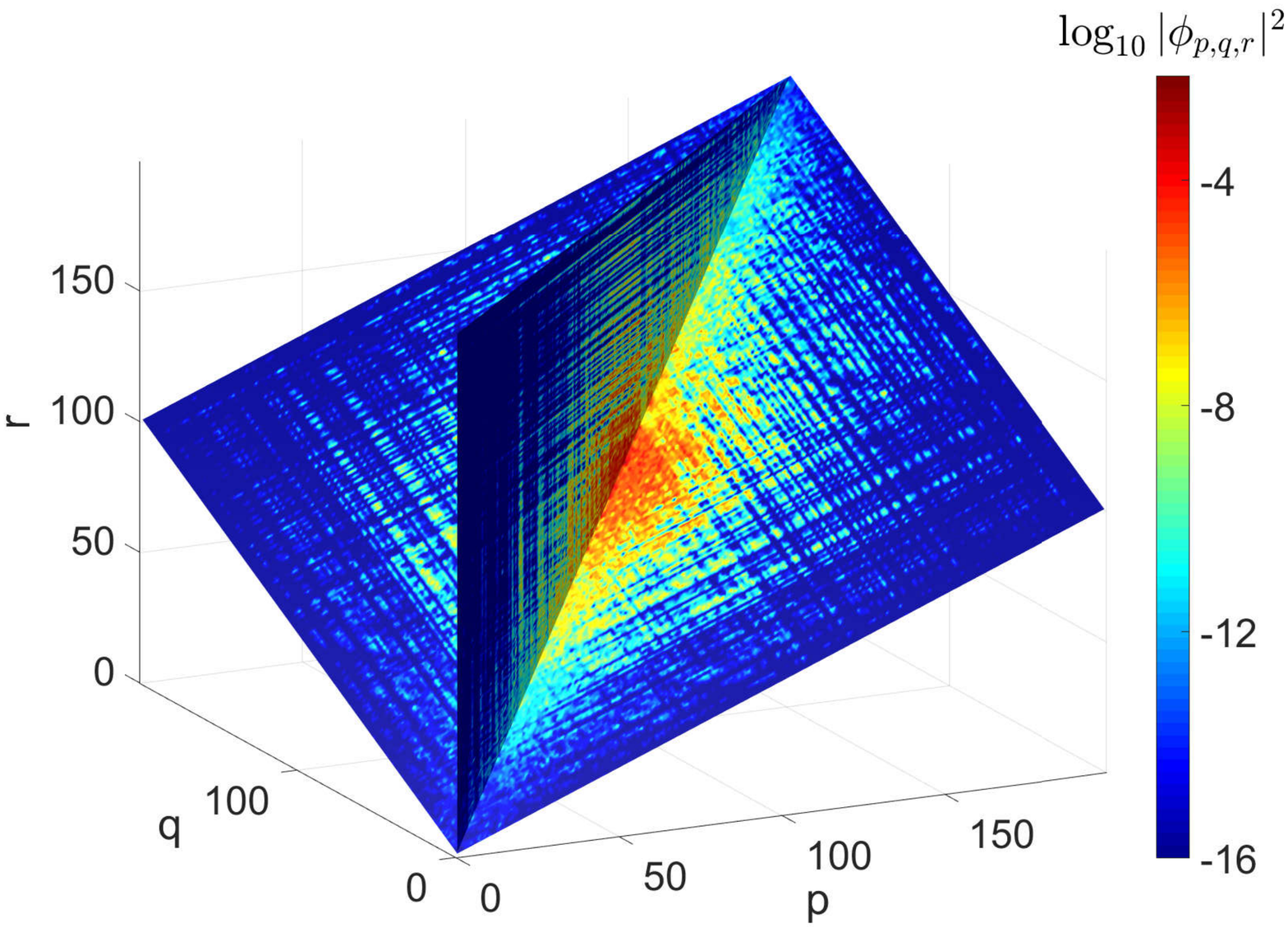}
(d)\includegraphics[angle=-90,width=0.9\columnwidth,keepaspectratio,clip]{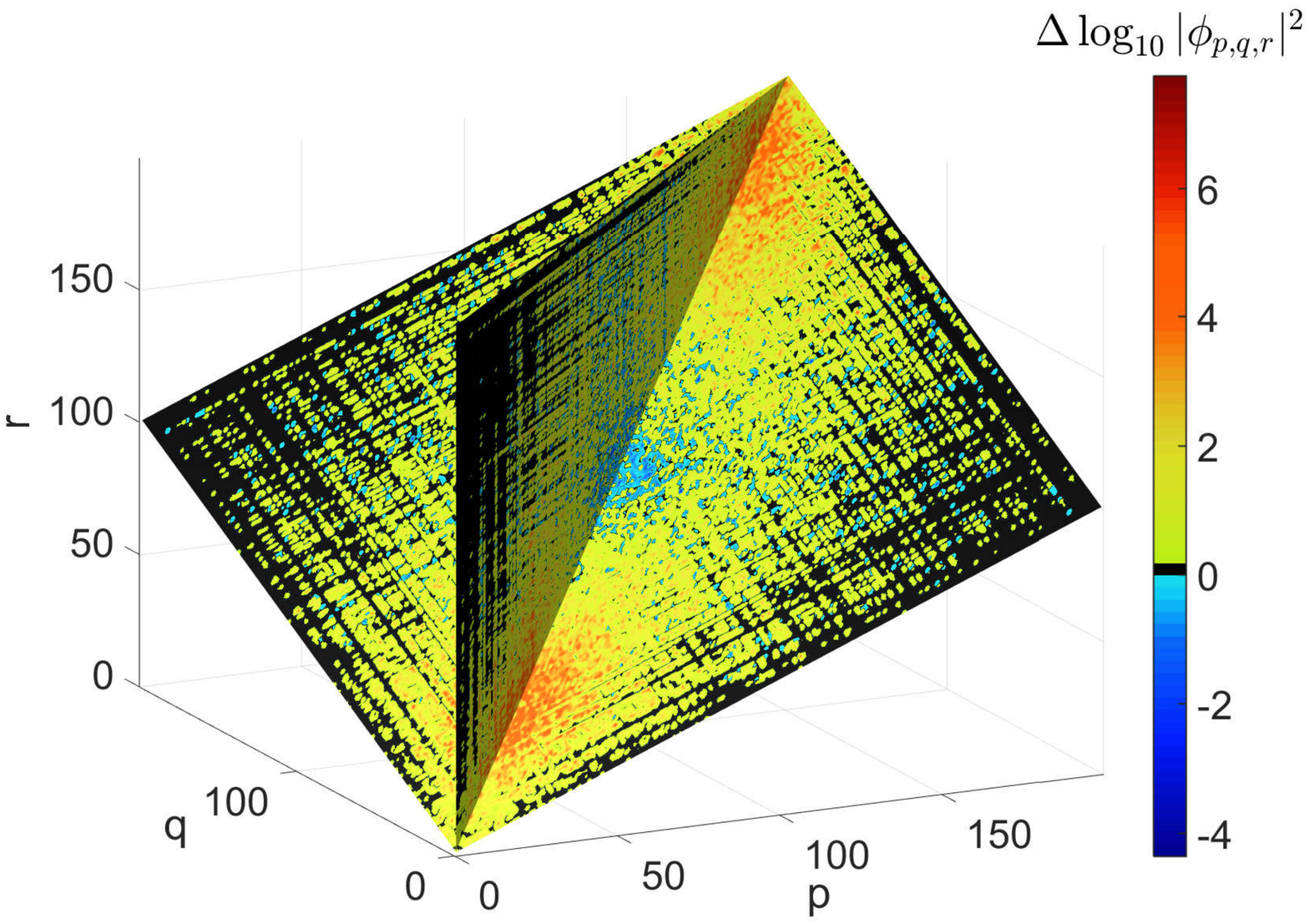}
\caption{Wave packet density distribution for a single Fock state initial condition, $\phi_{N/2,N/2}=1$ for 2IP, and $\phi_{N/2,N/2,N/2}=1$ for 3IP (see main text for details).
(a) 2IP, $W=2$, $U_2=2$, $N=1000$, $t=10^4$; (b) parameters as in (a) but now the difference between the distributions at $t=10^4$ and $t=10^3$ is plotted, indicating the diffusion of norm; 
(c) $\mbox{3IP}_3$, $W=3$, $U_2=0$, $U_3=2$, $N=200$, $t=10^3$; (d) parameters as in (a) but now the difference between distributions at $t=10^3$ and $t=10^2$ is plotted, indicating the diffusion of norm.
For (c) and (d) the plotted sections are $q=p$ and $p=r$, and their intersection corresponds to the main diagonal $q=p=r$. {Results for 2IP corroborate the earlier findings for smaller timescales in \cite{Ivanchenko2014}}. 
}
\label{fig:4}
\end{center}
\end{figure*}

Let us study the same wave packet dynamics using the Fock basis, constituted by  the eigenstates of the non-interacting system. For 2IP we use the transformation
 $\varphi_{j,k}=\sum_{p,q} \phi_{p,q} A_{j}^{(p)}A_{k}^{(q)}$ and arrive at
\begin{equation}
\label{eq4}
i\dot{\phi}_{p,q}=\lambda_{p,q}\phi_{p,q}+U_2\sum_{p',q'} I_{p,q,p',q'}\phi_{p',q'},
\end{equation} 
where $\lambda_{q,p}=\lambda_{p}+\lambda_{q}$ are the effective on-site energies and $I_{p,q,p',q'}=\sum_j A_j^{(p)}A_j^{(q)}A_j^{(p')}A_j^{(q')}$ are the (2nd order) overlap integrals, or simply the off-diagonal matrix elements
which provide with the corresponding hopping strength on the considered Fock network.
For 3IP we use $\varphi_{j,k,m}=\sum_{p,q,r} \phi_{p,q,r} A_{j}^{(p)}A_{j}^{(q)}A_{m}^{(r)}$ and arrive at
\begin{equation}
\label{eq5}
\begin{aligned}
&i\dot{\phi}_{p,q,r}=\lambda_{p,q,r}\phi_{p,q,r}+U_2\left(\sum_{p',q'} I_{p,q,p',q'}\phi_{p',q',r}\right.+\\
&\left.\sum_{p',r'} I_{p,r,p',r'}\phi_{p',q,r'}+\sum_{q',r'} I_{q,r,q',r'}\phi_{q',r'}\right)+\\
&U_3\sum_{p',q',r'} J_{p,q,r,p',q',r'}\phi_{p',q',r'},
\end{aligned}
\end{equation}
where $\lambda_{p,q,r}=\lambda_{p}+\lambda_{q}+\lambda_{r}$ and the 3d order overlap integrals are $J_{p,q,r,p',q',r'}=\sum_j A_j^{(p)}A_j^{(q)}A_j^{(r)}A_j^{(p')}A_j^{(q')}A_j^{(r')}$.
Note that for simplicity we added both two-body and three-body interactions into Eq.(\ref{eq5}), while all our results hold exclusively for either $U_2=0$ or $U_3=0$.

In this basis, particle interaction results in nonzero matrix elements (the hopping integrals) which connect different states in Fock space. 
Since all single particle states are Anderson localized in real space, we can sort them with increasing
mass center $j_c^{(p)}=\sum _j j A_j^{(p)}$. Ordering the modes by $j_c^{(p)}$, one obtains again effective two-dimensional (2IP) and  three-dimensional (3IP$_{2,3}$) lattices. Initial conditions localized in the original basis are localized there as well, and wave packet expansion on the original lattice necessarily involves excitation of new modes and spreading in the mode basis too. 

 A simple perturbative argument predicts strong interaction between the states, for which the resonance conditions {  $|\Delta\lambda| \lesssim |I,J|$ hold, where $\Delta\lambda=\lambda_{p,q}-\lambda_{p',q'}$ or $\Delta\lambda=\lambda_{p,q,r}-\lambda_{p',q',r'}$} \cite{Krimer2015}. As the overlap integrals decay exponentially with distance between the modes, $I,J\sim \exp(-|p-p'|/\xi_1)$, an effective interaction between modes is restricted to a distance less than $\xi_1$.
It follows, that the wave packet spreading has to be confined to the neighborhood of the diagonal of the Fock space lattice, such that positions of excited modes remain close, with additional selection by proximity of the Fock state energies.

This conclusion is confirmed by direct numerical simulations that depict the norm distribution in Fock space and parameters $W=2$, $N=1000$ after $t=10^4$ evolution time for 2IP, and
$W=3$, $N=200$ after $t=10^3$ evolution time for 3IP.
We start the dynamics with exciting
a single Fock state in the center of the Fock lattice, $\phi_{N/2,N/2}=1$ for 2IP, and $\phi_{N/2,N/2,N/2}=1$ for 3IP \cite{energy}. 
To begin with 2IP case, propagation along the diagonal, associated with interactions between $p\approx q$ and $p'\approx q'$ Fock states, is accompanied by a rich structure of vertical and horizontal walks (Fig.\ref{fig:4}(a)). The latter are produced by the near-resonant interaction between the Fock states, for which one of the indexes coincides, $(p,q)$ and $(p',q)$. It requires, in particular, $|\lambda_p-\lambda_p'|\lesssim I_{p,q,p',q}$, which {  can be realized at least for some of the modes from a single-particle localization volume, $|p-p'|\lesssim\xi_1$, whose overlap integrals are large \cite{Pichard1998,Krimer2015}.} {  Beyond $\xi_1$ the overlap integrals decay exponentially and further wave packet propagation in vertical and horizontal directions} is suppressed. Finding an alternative near resonance away from the vertical or horizontal line is even less probable, as $q'\neq q$ {  immediately boosts the energy mismatch between the Fock states, $\Delta\lambda$ (recall that the spatial ordering of the modes in Fig.\ref{fig:4} does not maintain continuity in energy), in addition to decaying overlap integrals.} 

Plotting the difference between the norm distributions in the Fock space at different moments of time, one clearly distinguishes the two main types of propagation.
 The first one takes place along the diagonal, $q\approx p$, and the second one along a single coordinate, for selected $p$ or $q$.
(Fig.\ref{fig:4}(b)). These observations are also valid in the case of 3IP, $W=3.0, U_2=0.0, U_3=2.0, N=200$ at $t=10^3$ (Fig.\ref{fig:4}(c,d)), where propagation is again taking place along the
main diagonal, $q\approx p \approx r$, and, simultaneously, along a single coordinate, for some fixed $p=q$, $q=r$ or $p=r$.

This renders a close resemblance to the diffusion on comb lattices \cite{comb1}, the backbone presented by the diagonal, supported by the near-resonances between the quadruplets, $|\lambda_p+\lambda_q|\approx|\lambda_{p'}+\lambda_{q'}|$, along with the horizontal and vertical `fingers', supported by the pairwise near-resonances, $\lambda_p\approx\lambda_{p'}$, with no significant diffusion between the `fingers'. It is well-known that such systems develop subdiffusion in the backbone direction due to excursions of particles along the fingers \cite{comb1,comb2,comb3}. 
The effects of the order of interaction can be understood as the effects of an effective dimension increase. Indeed, in the case $U_2\neq0$, diffusion of 3IP is dominated by propagating pairs.
the diffusion in Fock space becomes effectively two-dimensional, resulting in the same exponent, as in the 2IP case. In contrast, when two-body interactions are forbidden, $U_2=0$ and $U_3\neq0$, 
diffusion develops in the full three-dimensional Fock space, along the diagonal as a back bone, and `fingers' possible in the two other dimensions. The increased number of potential directions for excursions away from the backbone of the multi-dimensional comb lattice slows the diffusion along the backbone direction even further. Indeed, the subdiffusion exponent decreases with the increase of the dimension of the comb lattice, $d$, as $\alpha\sim 1/2^{d-1}$ \cite{comb3}. In particular, it yields $\alpha=1/2$ in two dimensions and $\alpha=1/4$ in three dimensions, in an intriguing correspondence to our numerical observations.

The mean field limit for many interacting particles yields exponents which also depend on the order of the many-body interactions. In particular it follows that $\alpha=1/2$ for two-body interactions and
$\alpha=1/3$ for three-body interactions in the so-called strong chaos regime \cite{Flach09,nonlinearity_HO} (these numbers change to $\alpha=1/3$ and $\alpha=1/5$ in the weak chaos regime).
This is in semi-quantitative agreement with our presented data on few body quantum interactions. However the mean field limit yields density distributions which are remarkably different from the 
ones observed in the present work. The mean field wave function develops a thermalized core with almost constant density, bounded by exponentially localized tails with time independent slopes corresponding to $\xi_1$ and
being independent of time. Spreading taking place there through a growth of the core width. This is very different to the quantum few-body case, where we report on time dependent slopes of the exponentially
localized distribution, and no evidence of a widening core with constant density.


\section{Conclusion}

We studied the dynamics of few interacting particles in the weak localization regime on a disordered one-dimensional lattice and observed subdiffusive spreading beyond the single particle localization length. The subdiffusion exponent is determined by the type of many-body interactions: two-body interactions yield $\alpha\approx0.5$ for both 2IP and 3IP, while three-body interactions reduce this number to 
$\alpha\approx0.2$. We relate our findings by comparing wave packet diffusion in the Fock basis with diffusion on comb lattices. These comparisons leave us with a number of open questions,
which will be addressed in future studies.
The dynamics of ultracold atomic and polar molecular condensates in modulated potentials is a promising test bed field for our findings. 

{\it Acknowledgments:} This work was supported by the Russian Science Foundation grant No.\ 15-12-20029 (I.Y., T.L., A.P., I.M., M.I.), and by the Institute for Basic Science (Project Code IBS-R024-D1, S.F.). I.Y., T.L., A.P. and I.M. performed numerical experiments and analyzed the results, M.I. and S.F. designed the study, analyzed the results and wrote the paper.



\begin{thebibliography}{99}
\bibitem{energy} The eigenenergy of the corresponding single particle state  varies with the disorder realization, since the energy of the state does not correlate with its center of mass position. The observed spreading dynamics in the Fock space does not change however qualitatively when different initial states are chosen.
\end{thebibliography}
\end{document}